# Identifying the determinants of soccer coach turnover by competing risks models: a case study on Italian League Serie A

Francesco Porro[1*], Marialuisa Restaino[2], Juan Eloy Ruiz-Castro[3] and Mariangela Zenga[4*]

*Correspondence:
Francesco Porro
francesco.porro@unige.it
Mariangela Zenga
mariangela.zenga@unimib.it
[1]Department of Mathematics, University of Genoa, Genoa, Italy
[2]Department of Economics and Statistics, University of Salerno, Fisciano (SA), Italy
[3]Department of Statistics and Operations Research and IMAG, University of Granada, Granada, Spain
[4]Department of Statistics and Quantitative Methods, University of Milano-Bicocca, Milan, Italy

**Abstract**
This paper investigates the determinants of head coach turnover in the Italian Serie A over the period 2010–2019 by explicitly accounting for multiple exit mechanisms within a competing risks framework. Four mutually exclusive exit types are considered: end of contract, mutual consent, resignation, and sacking. To identify robust predictors of turnover, we combine classical cause-specific Cox models with a variable selection procedure based on forward selection, LASSO regression, and Random Survival Forests. The results show that performance-related variables are the primary drivers of coach turnover, while personal characteristics play a limited role once performance is controlled for. Short-term performance indicators, such as points per match until turnover, strongly influence the risk of dismissal. Conversely, cumulative poor performance substantially increases dismissal risk: a higher proportion of losses over a coach's tenure raises the hazard of sacking by more than four times. Season timing also plays a crucial role, as sackings and mutual-consent terminations are significantly more likely to occur during the season, whereas contract expirations concentrate at the end of the season. Overall, the findings highlight that clubs differentiate between short-term performance shocks and longer-term performance trajectories when deciding how and when to replace a head coach. By distinguishing among exit mechanisms and adopting a conservative variable selection strategy, the study provides a robust and nuanced assessment of managerial turnover dynamics in the most important Italian professional football league.

**Keywords** Coach turnover, Random survival forest, Competing risks model, Variable selection, Forward Selection, LASSO, Survival analysis, Sports analytics

## Introduction

Recent decades have seen a significant increase in interest in sports, team performance, and, in particular, the careers of head coaches. The two primary drivers of this interest are the structural similarities between the labor markets and sports teams and the increasing availability of granular datasets spanning numerous teams and sports [1]. Several studies have demonstrated these similarities [2, 3], enabling the analysis of specific sports labor markets and the subsequent extension of empirical findings to the broader labor market.

In the sporting context, the study of head coaches' careers is crucial for understanding both team performance and financial stability, as hiring and firing decisions significantly affect the team's success. Analyzing the likelihood of coach turnover provides insights into team management dynamics and offers a framework for linking performance, managerial characteristics, and organizational behavior. Moreover, head coaches can be compared to Chief Executive Officers (CEOs) in traditional firms; like soccer club managers, they are appointed to improve organizational performance [4].

Head coaches oversee a wide range of responsibilities, including recruiting players, selecting matches, managing support staff, and defining tactical strategies to defeat opponents. Since head coaches serve as team leaders and play a pivotal role in performance, their dismissal and the likelihood of turnover are highly relevant to several stakeholders, including fans, sponsors, team owners, and players. Football clubs can use turnover studies to improve hiring practices and mitigate the risk of losing key personnel. At the same time, current and prospective coaches may gain a better understanding of the factors influencing job stability. Additionally, experts and fans can evaluate coaching tenures within an evidence-based framework, while investors can use such insights to guide decisions regarding managerial stability.

Given its relevance to such stakeholders, the role of a head coach mirrors that of a corporate CEO: when performance is poor, replacing the leader is often viewed as a primary corrective measure. However, terminating a contract before its natural expiration can disrupt decision-making and lead to significant costs [5, 6]. Consequently, a rich body of literature has investigated the effects of managerial replacements in professional sports–primarily soccer–across several European countries (for an overview, see [7]).

From an organizational and behavioral perspective, managerial turnover is commonly interpreted as the outcome of heterogeneous decision-making processes driven by performance evaluation, contractual arrangements, and individual incentives. A foundational distinction in the turnover literature exists between *voluntary turnover*, arising from an individual's decision to leave, and *involuntary turnover*, resulting from employer-initiated dismissals or contract non-renewals [8, 9]. These mechanisms are associated with distinct behavioral responses to performance signals, job satisfaction, external opportunities, and organizational pressure [8, 9].

In the context of professional soccer, these distinctions translate naturally into different exit modes for head coaches, such as resignation, dismissal, or contract expiration. Each exit type reflects a distinct underlying behavioral and organizational process, suggesting they should not be treated as observationally equivalent outcomes [2, 10]. As emphasized by Kahn [2], the transparency of performance outcomes in sports allows managerial decisions to be analyzed in ways that are often difficult in other organizational contexts. The multiplicity of contract outcomes and exit modes observed in professional sports further reinforces the suitability of a competing risks framework for modeling managerial turnover in this environment.

Although research on the termination of head coach contracts is active [11–14], no universally accepted model has yet emerged. Several studies have relied on duration and survival analysis frameworks to investigate coach turnover [7, 15–17]. Nevertheless, there is no consensus regarding which characteristics of coaches, teams, or clubs most influence the risk of dismissal.

Building on these considerations, this paper's contributions to the debate on coach turnover are threefold.

First, we construct an extensive database (comprising over 50 variables) detailing the characteristics of coaches, teams, and clubs influencing head coach replacements in the Italian top football division (Serie A) over the 2010/11–2019/20 seasons. The focus on Serie A is motivated by its status as one of the world's "Big Five" leagues, its historical importance, and its notably high rate of coach turnover. Frequent management replacements make Serie A an ideal setting for analyzing the impact of such decisions. Moreover, as Serie A is a multibillion-euro industry, managerial continuity plays a crucial role in long-term success [18]. Although the analysis focuses on Italy, the findings provide a useful starting point for comparative studies of other top leagues and high-performance sectors, where decision-making under risk is similarly critical.

Second, this paper analyzes the characteristics of coach turnover using a competing risks model, an extension of classical survival analysis [7, 15–17]. This model allows us to classify coach exits into four distinct types: contract expiration without renewal, termination by mutual consent, sacking, and resignation. To the best of our knowledge, no existing studies have applied competing risks models in this context. The competing risks model, therefore, provides a more nuanced tool for understanding the mechanisms underlying turnover.

Third, given the large number of potential predictors, we explicitly address variable selection within the competing risks framework. Variable selection is a crucial issue in survival analysis when the number of potential predictors is large, as it improves interpretability and predictive accuracy while reducing the risk of overfitting [19, 20]. To this end, we implement a multi-method variable selection strategy based on three well-established procedures: forward selection, a classical stepwise procedure used as a benchmark in regression models [21, 22]; penalized regression via the Least Absolute Shrinkage and Selection Operator (LASSO), a standard technique for high-dimensional Cox and competing risks models [19, 23]; and random survival forests, a nonparametric machine-learning method that extends Breiman's random forests [24] to right-censored data, making it particularly suitable for complex, nonlinear relationships and Big Data settings [20, 25, 26]. By combining traditional, penalized, and nonparametric approaches, we can robustly identify the most relevant predictors of turnover within both classical and modern data-driven frameworks.

Given these objectives, our framework is motivated by two methodological considerations. First, coach turnover involves multiple mutually exclusive exit states. Treating these outcomes as a single event, or collapsing them into a binary indicator (as in probit or logit models), implicitly assumes homogeneity across fundamentally different exit mechanisms. Such an assumption may lead to inconsistent estimates if covariates affect competing events in opposite directions. The competing risks framework accounts for this heterogeneity by modeling cause-specific hazard functions. Second, our empirical analysis relies on a large set of potentially correlated covariates. In this context, standard Cox models may suffer from overfitting and instability. To address the issue, our variable selection strategy ensures robustness: forward selection facilitates comparability with existing studies, LASSO introduces regularization to enhance stability, and random survival forests capture nonlinearities and complex interactions. The combined approach reduces the risk that results are driven by a specific modeling assumption.

Thus, by comparing parametric, semi-parametric, and nonparametric approaches within a unified competing risks framework, our study contributes to the growing literature on data-driven survival analysis in applied contexts.

The remainder of this paper is organized as follows. Section "Literature review" provides an overview of the related literature and the methodological approaches used in studies on coach turnover. Section "Methodological framework" describes the statistical methods: Sect. "Competing risks model for the coach turnover" introduces the competing risks framework and its features, Sect. "Variable selection strategy in the competing risks model" illustrates the variable selection strategy implemented in the study, and Sect. "The three variable selection methods" details the corresponding procedures. Section "Data and variable construction" describes the dataset and the variables. Section "Empirical results" is devoted to the empirical findings: Sect. "Relevant covariates by exit mechanism" presents the results of the variable selection procedure, while Sect. "Cause-specific hazard ratios by exit mechanism" provides hazard ratio estimates and analyzes the selected predictors. Finally, Sect. "Discussion and conclusions" summarizes the main findings and discusses their broader implications, with Sect. "Managerial and practical implications" focusing on the managerial implications.

## Literature review

In the applied econometric and statistical literature, professional sports have increasingly been used as a laboratory for analyzing labor market dynamics, given the availability of detailed performance and contract data. Several methods from diverse fields have been employed for various analytical purposes. For a comprehensive review of the recent empirical studies and methodological contributions, see [1] and the references therein.

Despite the growing methodological sophistication of this literature, however, there is still no consensus on a unified framework capable of jointly handling multiple exit types and high-dimensional covariate spaces.

Beyond the prediction of match outcomes (see [27, 28], and [29] for recent findings related to on-field variables), other topics are attracting considerable attention. Among them, the analysis of which factors can help the study and the forecast of the firing or dismissal of a team's head coach has been widely explored. In many cases, the role of a head coach can be compared to that of a corporate Chief Executive Officer (CEO).

Beyond sport-specific studies, a large body of research in organizational and labor economics has examined managerial turnover through the lenses of performance evaluation, incentive alignment, and accountability. Empirical studies on CEOs and top managers consistently document a strong sensitivity of involuntary turnover to firm performance. At the same time, voluntary departures are more closely associated with career concerns, external opportunities, and contractual conditions [30–32]. For example, early studies show that poor firm performance significantly increases the likelihood of CEO dismissal [30]. Furthermore, agency-theoretic models emphasize the role of performance-based incentives and accountability mechanisms in disciplining managers [31]. More recent evidence indicates that CEO turnover has become increasingly performance-sensitive over time, reflecting stronger governance structures and monitoring practices [32]. Professional sports represent a particularly suitable environment for testing these theories, as managerial performance is observed at high frequency and outcomes are highly transparent. In this sense, head coach turnover closely mirrors managerial turnover in firms, with clubs acting as principals who evaluate agents based on short-term and cumulative performance indicators.

Moreover, as observed by [15, 33, 34], and [35], applications in this field are compelling because performance measures for soccer teams can be readily derived from the large amounts of available data. This granularity allows for more precise analyses of performance dynamics than are typically feasible in studies of general firm performance. In the literature, analyses of the reasons for head coach dismissal have been reported using several methodologies. Restricting attention to soccer, a list of recent relevant papers is presented in Table 1.

The studies by [13, 18], and [36] use probit models, while [14] and [38] apply logistic regression to analyze major European leagues, including Spain, Italy, and Germany. [39] propose a standard logistic model that includes covariates related to team performance and the coach's player experience, finding that league position is the most relevant determinant of a manager's status.

A common feature of the studies summarized in Table 1 is their reliance on binary-choice models or single-exit duration models, which implicitly treat coach turnover as a homogeneous event [14, 41, 42]. Such approaches do not distinguish among exit mechanisms and therefore overlook the possibility that voluntary and involuntary turnover may respond differently to performance and managerial characteristics [8, 9, 30]. Moreover, while these approaches have provided valuable insights, they typically abstract from the fact that coach turnover may occur through qualitatively distinct mechanisms that are treated as observationally equivalent [7, 10]. From a theoretical standpoint, this aggregation may obscure heterogeneous covariate effects and lead to biased inference when the risk structure differs across exit types. Although survival analysis has been increasingly adopted, most studies either focus on a single failure type or do not explicitly address variable selection in the presence of a rich set of regressors [14, 43]. In addition, several contributions include only a limited number of covariates selected a priori, which may lead to omitted-variable bias when rich datasets are available [44]. These limitations motivate the adoption of a competing risks framework combined with multiple variable selection techniques, enabling a more flexible, data-driven, and robust identification of the determinants of coach turnover.

**Table 1** Recent papers on the dismissal of soccer head coaches

| Author(s) and year | Methodology | Variable selection |
|---|---|---|
| Bryson et al. 2021 [4] | Survival Analysis | Yes |
| Tena 2007 [13] | Ordered Probit Model | No |
| De Paola and Scoppa 2012 [18] | Probit Model, Fixed Effect Model | No |
| Detotto et al. 2018 [36] | Ordered Probit Model | Yes |
| Frick et al. 2010 [14] | Mixed Logistic Model | No |
| Gilflix et al. 2020 [16] | Survival Analysis | No |
| Tozetto et al. 2019 [17] | Multilevel Survival Analysis | No |
| Zambom-Ferraresi et al. 2019 [37] | Data Envelopment Analysis | No |
| van Ours and van Tuijl 2016 [7] | Survival Analysis | No |
| Daddona and Kind 2014 [15] | Duration Analysis | No |
| Salomo and Teichmann 2000 [38] | Logistic Regression | No |
| Bachan et al. 2008 [39] | Logistic Regression | Yes |
| Barros et al. 2009 [40] | Duration model | No |
| Pieper et al. 2014 [6] | Linear Probability Model | No |

An empirical investigation of how performance expectations affect the replacement or retention of top managers is described in [6]. The authors, using bookmakers' betting odds as a proxy for performance expectations, employ a linear probability model to assess their effect on both involuntary and voluntary turnover. Moreover, [37] apply a bootstrapped Data Envelopment Analysis (DEA) approach to compare ex-ante inputs (e.g., market value) with ex-post inputs (e.g., match statistics): their findings indicate a positive, significant correlation between the two models.

Another relevant methodology in quantitative sports research is survival analysis. [15] study the English Premier League using duration analysis, considering performance measures alongside coach characteristics such as age and experience. Similarly, the model by [40] includes coach and team salaries as explanatory variables. More advanced competing risks models are employed in [4], who demonstrate that the probability of dismissal decreases when a team performs above expectations or when the coach is highly experienced, suggesting a "protective" effect. Conversely, prior experience has little impact on the probability of resignation.

In [7], two related issues are examined: the determinants of in-season dismissal and the effects of a coach's departure on subsequent performance. Their findings suggest that while coach experience and performance indicators (such as points and betting odds) significantly impact dismissal rates, replacing a coach does not necessarily improve subsequent team performance. Survival analytic methods, including Cox's proportional hazards model, are also applied by [16] to Major League Soccer (United States), La Liga (Spain), and the Premier League (United Kingdom). Their analysis shows that, while the tenure for domestic coaches in the United States is significantly longer than that of foreigners, such difference is minimal in European leagues.

[17] introduce a multilevel hierarchical survival model applied to the Brazilian Série A. Their findings indicate that prior experience in the top division does not significantly affect the survival probability, highlighting the vulnerability of coaches in environments where short-term results heavily drive club management decisions.

From a theoretical standpoint, the presence of multiple exit modes implies that a single failure event cannot adequately represent coach turnover. Voluntary and involuntary exits correspond to distinct behavioral mechanisms and organizational decisions; covariates such as performance, season timing, and contractual history may affect these mechanisms differently [9, 32]. Treating all exits as a single outcome may therefore obscure essential sources of heterogeneity and may lead to misleading inference.

From a methodological perspective, no single variable selection technique is fully suited to address the combination of challenges posed by coach turnover data, including competing risks, high-dimensional covariates, non-linear effects, and the need for interpretable cause-specific inference. Tree-based methods, such as Random Survival Forests (RSF), are well-suited to capturing complex interactions and non-linearities without strong parametric assumptions, but they are not designed for direct inferential interpretation. Conversely, classical Cox-type models allow for transparent interpretation but may suffer from overfitting in high-dimensional settings. The integration of RSF, forward selection in cause-specific Cox models, and LASSO regularization is therefore motivated by their complementary strengths. RSF serves as an exploratory screening tool, forward selection provides an interpretable econometric benchmark, and LASSO addresses sparsity and multicollinearity. This multi-method approach reflects a principled strategy for balancing flexibility, interpretability, and statistical robustness. Our approach is explicitly tailored to the competing risks setting to disentangle the determinants of distinct exit modes.

The competing risks framework provides a natural statistical representation of these considerations by modeling each exit type as a separate but mutually exclusive risk [45, 46]. This approach is consistent with organizational theories of accountability, which recognize that different processes govern distinct forms of managerial exit. Several recent studies have adopted hybrid strategies combining tree-based screening (e.g., RSF) with penalized regression for final inference. In biomedical and engineering applications, researchers commonly use RSF to identify relevant predictors and then apply penalized Cox models (LASSO/Elastic Net) for interpretable estimates [47–49]. Our multi-method strategy follows this pragmatic tradition, adapting it to the competing-risks context to preserve cause-specific interpretability while leveraging RSF's robustness to nonlinearities. This hybrid approach aligns with recent advances in survival analysis, which emphasize that no single method is uniformly optimal [47, 50, 51].

In summary, although the existing literature provides important insights into coach turnover, it is largely constrained by modeling strategies that either collapse heterogeneous exit mechanisms or rely on a limited set of covariates selected a priori. By contrast, our approach jointly addresses both issues by combining a competing risks framework–consistent with organizational theories of turnover–with modern variable selection techniques suitable for high-dimensional data. This enables a more accurate identification of the determinants of distinct exit modes, mitigating the risks of overfitting and omitted-variable bias.

## Methodological framework

We examine head coach turnover by estimating the probability of coach replacement using survival techniques. In particular, we introduce a semi-parametric model to analyze survival data while considering multiple events of interest, i.e., a competing risks model (Sect. "Competing risks model for the coach turnover"). The use of competing risks is motivated by the fact that head coach turnover may occur through qualitatively different exit mechanisms–such as dismissals, resignations, or contract expirations–which reflect distinct organizational and behavioral processes and should therefore not be treated as observationally equivalent.

Since the number of features affecting the probability of coach turnover is likely high, the most relevant ones are selected using three variable selection methods: forward selection [21], LASSO regression [19], and the nonparametric technique random survival forest (RSF) [26].

The combined use of classical and machine-learning-based selection procedures allows us to balance interpretability and flexibility, while assessing the robustness of selected covariates across methods that rely on fundamentally different modeling assumptions.

### Competing risks model for the coach turnover

In this section, we introduce notation and recall the main elements of the theoretical framework used in this study to model the probability of head-coach turnover.

As multiple types of coaching replacements may occur, reflecting distinct organizational and contractual processes, as discussed in Sects. "Introduction" and "Literature review", we adopt a competing risks framework [52, 53] that extends standard survival models by explicitly accounting for multiple and mutually exclusive events of interest.

Distinguishing among exit mechanisms is therefore crucial for interpreting turnover as an outcome of organizational decision-making rather than as a purely stochastic event.

Let $(\Omega, \mathcal{F}, \mathbb{P})$ be a probability space. For each coach–team spell, we define the following random variables. Let $\tilde{T}$ denote the latent failure time, representing the unobserved time elapsed from the beginning of the season until the occurrence of a coach turnover event. Turnover is defined as the termination of a coaching spell due to one of the exit mechanisms considered in this study.

Let $C$ denote the censoring time, defined as the time elapsed from the beginning of the season until the end of the observation window. Right-censoring occurs when no turnover event is observed before the end of the season. We assume non-informative right-censoring, that is, $\tilde{T}$ and $C$ are independent conditional on the observed covariates.

The observed time is defined as

$$T = \min(\tilde{T}, C),$$

and the censoring indicator is given by

$$\delta = \mathbb{I}(\tilde{T} \leq C),$$

where $\mathbb{I}(\cdot)$ denotes the indicator function. The variable $\delta$ takes the value one when a turnover event is observed and zero otherwise, corresponding to right-censoring.

In addition, let $D$ be a discrete random variable identifying the cause of failure associated with a turnover event, taking values in the finite set $\{1, \ldots, K\}$, where $K = 4$ in our application. The competing events correspond to: (1) end of contract without renewal, (2) termination by mutual consent, (3) resignation, and (4) sacking. A detailed description of these exit types is provided in Section "Data and variable construction". The variable $D$ is observed only when $\delta = 1$; when $\delta = 0$, the failure cause is undefined.

Time is measured in days from the beginning of the season. Over the seasons from 2010/11 to 2019/20, we observe 337 coach–team spells with an observed turnover event, while spells without turnover by the end of the season are treated as right-censored observations.

The competing risks framework characterizes the joint distribution of $(\tilde{T}, D)$. Two main modeling approaches have been proposed in the literature: cause-specific hazard models and subdistribution hazard models [45]. In this study, we focus on the cause-specific hazard function, which is particularly appropriate when the objective is to assess how covariates affect the instantaneous risk of a specific exit type, rather than to model cumulative incidence. This choice is also motivated by the fact that our primary interest lies in understanding the determinants of distinct turnover mechanisms, rather than in forecasting cumulative incidence probabilities.

Formally, the cause-specific hazard function for event type $k$ [54] is defined as

$$\lambda_k^{cs}(t) = \lim_{\Delta t \to 0} \frac{\mathbb{P}(t < T \leq t + \Delta t, D = k \mid T > t)}{\Delta t}, \quad k = 1, \ldots, K. \tag{1}$$

In Formula 1, the expression $\mathbb{P}(t < T \leq t + \Delta t | T > t)$ represents the standard definition of the hazard rate in survival analysis. While the condition $T > t$ logically implies $t < T$, the full notation is maintained to explicitly distinguish between the risk set (the condition) and the failure event occurring in the specific infinitesimal time interval $[t, t + \Delta t)$.

The cause-specific hazard represents the instantaneous risk of experiencing a turnover of type *k* at time *t*, conditional on the coach having remained in the position up to time *t*.

The subdistribution hazard function [52] instead captures the instantaneous risk of failure from cause *k*, given that the coach has not yet experienced an event of type *k*, and is defined as

$$\lambda_k^{sd}(t) = \lim_{\Delta t \to 0} \frac{\mathbb{P}(t < T \leq t + \Delta t, D = k \mid T > t \cup (T \leq t, D \neq k))}{\Delta t}, \quad k = 1, \ldots, K. \tag{2}$$

Regarding Formula 2, the conditioning set $T > t \cup (T \leq t, D \neq k)$ is not redundant but is the defining characteristic of the subdistribution hazard. This formulation intentionally keeps individuals who have experienced a competing event ($D \neq k$) in the risk set. This distinct structure is what allows the subdistribution hazard to be directly related to the cumulative incidence function, differentiating it from the cause-specific hazard in Formula (1) where the risk set includes those who are event-free only.

The key distinction between the cause-specific hazard in Eq. (1) and the subdistribution hazard in Eq. (2) lies in the definition of the risk set. For the cause-specific hazard, the risk set at time *t* consists of all coaches who have not experienced any turnover event prior to *t*. In contrast, the subdistribution hazard augments the risk set for cause *k* by retaining coaches who have previously experienced a turnover due to causes other than *k*, even though they can no longer experience event *k*.

Consequently, under the cause-specific hazard function, the risk set shrinks whenever a turnover occurs due to any competing cause. By contrast, under the subdistribution hazard approach, coaches who fail from competing causes remain in the risk set and are assigned censoring times beyond all observed event times.

In this study, each coach–team spell is subject to at most one turnover event, and the different exit types are mutually exclusive. The realization of one event type, therefore, precludes the occurrence of all remaining events. For this reason, we focus exclusively on the cause-specific hazard function in the empirical analysis.

Let $\mathrm{Z}_k(t) \in \mathbb{R}^p$ denote a *p*-dimensional vector of possibly time-varying covariates associated with exit type *k*. The conditional cause-specific hazard function given covariates is defined as

$$\lambda_k^{cs}(t \mid \mathrm{Z}_k(t)) = \lim_{\Delta t \to 0} \frac{\mathbb{P}(T \leq t + \Delta t, D = k \mid T > t, \mathrm{Z}_k(t))}{\Delta t}, \quad k = 1, \ldots, K. \tag{3}$$

The subscript *k* in $\mathrm{Z}_k(t)$ indicates that covariate effects are allowed to differ across exit types, thereby capturing potentially heterogeneous organizational responses to performance and contextual factors. In the empirical application, the same set of observed regressors is included for all causes, while the associated coefficient vectors $\boldsymbol{\beta}_k$ are permitted to vary across exit types.

Moreover, in our analysis, all covariates included in the competing risks models are treated as time-independent $Z_k$. This choice reflects the structure of the available data and the analysis's focus on coach–team spells rather than on match-level dynamics. Moreover, this choice is consistent with prior survival-based studies on managerial turnover that operate at the spell level rather than the match level [7]. Performance variables are defined as cumulative or average measures computed over fixed horizons (e.g., until turnover, over the season, or over the entire tenure), and thus summarize information available to clubs at the time decisions are made. While survival models can accommodate time-varying covariates, incorporating match-by-match fluctuations would require a substantially different data structure and modeling framework. The adopted specification allows for a clear interpretation of covariate effects and is standard in the literature on managerial turnover.

Cause-specific Cox regression extends the standard Cox proportional hazards model to the competing risks setting by estimating a separate hazard for each exit type [45]. Accordingly, for each cause *k*, all coach–team spells experiencing a different exit type are treated as censored at the time of occurrence.

$$\lambda_k^{cs}(t \mid \mathrm{Z}_k(t)) = \lambda_{k,0}(t) \exp\{\mathrm{Z}_k^\top(t)\boldsymbol{\beta}_k\}, \tag{4}$$

where $\lambda_{k,0}(t)$ denotes the cause-specific baseline hazard function, which is left unspecified, $\mathrm{Z}_k(t)$ is a vector of $p$ covariates, and $\boldsymbol{\beta}_k$ is the corresponding vector of unknown coefficients for exit type *k*.

The cause-specific hazard function is particularly well suited to our research objective, as it allows us to examine how covariates affect the instantaneous risk of each specific exit type, rather than focusing on cumulative incidence probabilities. This perspective is consistent with organizational theories of accountability, which emphasize the timing and conditions under which managerial decisions are taken (e.g., [8, 9, 32]).

The coefficient vector $\boldsymbol{\beta}_k$ captures the effect of covariates on the instantaneous risk of experiencing a turnover of type *k*. For a given covariate, the quantity $\exp(\beta_{kj})$ represents the cause-specific hazard ratio associated with a one-unit increase in that covariate, conditional on survival up to time *t*.

To estimate the coefficient vectors $\boldsymbol{\beta}_k$, the partial likelihood function associated with each cause-specific hazard is maximized using standard Cox proportional hazards techniques [55].

To assess model adequacy and predictive performance, although the primary objective of this study is explanatory rather than purely predictive, we compute standard survival-analysis metrics. For the cause-specific Cox models, goodness-of-fit and discriminatory power are evaluated using the concordance index (C-index), which measures the model's ability to rank survival times correctly.

### Variable selection strategy in the competing risks model

The number of observable characteristics that may affect the probability of coach turnover is potentially high, while only a subset is expected to be truly relevant for each exit mechanism. Accordingly, we adopt a multi-method variable selection strategy that combines three methodologies, balancing flexibility, parsimony, and interpretability in the competing risks setting. Our strategy uses forward selection [21], LASSO regression [19, 23] and Random Survival Forests (RSF) [20, 25]. The idea is to include in the final model only the variables that resulted as relevant by all three methods. We adopted this approach for consistency. Indeed, such three-variable selection procedures are not used as competing alternatives but as complementary tools that address different statistical aspects of the same inferential problem, as no single method is expected to simultaneously capture nonlinear effects, control model complexity, and provide transparent inference in high-dimensional competing risks settings. Methodologically, our goal is to create a data-driven screening and ranking tool for uncovering nonlinearities and interactions without parametric assumptions using RSF; for accounting for sparsity and mitigating multicollinearity within cause-specific Cox models through LASSO; and for providing a transparent classical procedure via forward selection. The aim of this design is to mitigate the risk of model misspecification and preserve the interpretability of covariate effects across exit types in the competing risks framework.

## The three variable selection methods

### *Forward selection*

Forward selection is a stepwise variable selection procedure commonly used in regression and survival analysis [56]. The procedure starts with a null model that includes only the intercept. At each iteration, all remaining candidate covariates are individually added to the current model, and the covariate that provides the largest improvement in model fit is selected. Model fit is evaluated using standard criteria, such as the statistical significance of coefficients, the Akaike Information Criterion (AIC), or the Bayesian Information Criterion (BIC). The procedure continues until a predefined stopping rule is met, for instance, when no additional covariate yields a statistically significant improvement or when information criteria no longer decrease.

Forward selection has several drawbacks. First, it can yield biased estimates and predictions because small replacements in the data may lead to significant changes in the set of variables included in the model. Second, multicollinearity among variables can increase the variance of the estimator coefficients, leading to inaccurate parameter estimates and erroneous identification of relevant predictors [22, 57].

### *LASSO regression*

As a second variable selection method, we consider the penalized regression based on the Least Absolute Shrinkage and Selection Operator (LASSO) [23]. In the competing risks setting, LASSO is implemented by maximizing a penalized partial likelihood, where an $\ell_1$ penalty is imposed on the regression coefficients

$$\hat{\boldsymbol{\beta}}_k = \underset{\boldsymbol{\beta}_k}{\operatorname{argmax}} \left[ \log(PL(\boldsymbol{\beta}_k)) - \lambda \sum_{j=1}^{p} |\boldsymbol{\beta}_{kj}| \right],$$

where $PL(\boldsymbol{\beta}_k)$ is the partial likelihood associated with the model described in Formula (4) for cause $k$, $\boldsymbol{\beta}_k$ is the vector of unknown coefficients for $p$ features and cause $k$, $\lambda \geq 0$ is the hyper-parameter that controls the amount of shrinkage. If $\lambda = 0$, the standard unpenalized model for the competing risks model is obtained. The penalty parameter $\lambda$ controls the amount of shrinkage and is selected via $k$-fold cross-validation, using the partial likelihood as the validation criterion. For a given value of $\lambda$, covariates with coefficients shrunk exactly to zero are excluded from the model, while those with non-zero coefficients are retained. This procedure yields a parsimonious specification and improves estimation stability in the presence of high-dimensional and potentially correlated covariates.

LASSO-based variable selection relies on $K$-fold cross-validation to determine the optimal penalization parameter, ensuring that selected covariates are not driven by in-sample overfitting. More details on this method can be found in [19, 23].

### *Random survival forests*

Random Survival Forests (RSF) constitute the machine-learning-based variable selection strategy employed in this study. They extend Breiman's random forest algorithm [24] to right-censored survival data [26]. RSF is a fully nonparametric ensemble method that is particularly well suited to high-dimensional survival analysis, as it can automatically capture nonlinear effects and complex interactions among covariates [20, 25]. The application of RSF has several advantages. First, as it is entirely data-driven, RSF does not require any model assumptions. Second, RSF is suitable for exploratory data analysis because it seeks the model that best explains the data, even when survival data are limited. Finally, RSF performs well on high-dimensional data and is robust to multicollinearity and outliers [58].

RSF is an ensemble tree-based method in which survival trees are grown by recursively partitioning the covariate space. Starting from the root node containing all observations, each split is selected to maximize differences in survival outcomes between child nodes. Through recursive splitting, observations within terminal nodes become increasingly homogeneous with respect to their survival experience. The construction of competing risks trees closely follows that of standard survival trees, with the key difference that splits are based on cause-specific hazards and cumulative incidence functions (CIFs) rather than overall survival. The CIF for the event $k$ is defined as

$$F_k(t) = \mathbb{P}(T \leq t, D = k) = \int_0^t S(u)\,\lambda_k^{cs}(u)\,\mathrm{d}u, \tag{5}$$

where $\lambda_k^{cs}(\cdot)$ is the cause-specific hazard function due to the event $k$ in Eq. (1), and $S(u) = \exp\left(-\sum_{j=1}^{K} \int_0^u \lambda_j^{\mathrm{cs}}(s)\,\mathrm{d}s\right)$ denotes the overall survival function.

The splitting rule in competing risks trees is based on the maximization of the differences in cumulative incidence between child nodes. By applying this splitting rule, the competing risks tree partitions observations according to the competing events, yielding terminal nodes that are increasingly homogeneous with respect to their hazard or cumulative incidence functions.

RSF combines bootstrap aggregation with random feature selection. Each tree is grown on a bootstrap sample drawn from the learning dataset. In contrast, observations not included in the bootstrap sample (out-of-bag, OOB, observations) are used for internal validation and performance assessment. Tree growth continues until a minimum number of unique events is reached in each terminal node, ensuring stable estimation of survival and cumulative incidence functions [25].

Variable importance is assessed using event-specific variable importance measures (VIMP), computed by comparing prediction errors before and after random permutation of a given covariate. A high VIMP value indicates that the variable has predictive ability, while zero or negative values indicate nonpredictive variables. As an additional criterion, minimal depth is used to rank variables by their proximity to the trees' root nodes, with smaller depths indicating greater predictive relevance. While VIMP provides an event-specific, prediction-based measure of relevance, minimal depth offers a global ranking based on forest structure.

The prediction accuracy is assessed using out-of-bag (OOB) concordance index (C-index) error estimates, which provide an internal form of cross-validation.

More details of the RSF can be found in [25].

## Data and variable construction

### Data sources and preprocessing

The data used to analyze coach turnovers are related to the top Italian soccer league (Serie A), and in particular, refer to 103 head coaches who took charge of games played by 34 Italian teams in the seasons from 2010/11 to 2019/20.

The dataset in this study is constructed by collecting and integrating information from multiple publicly available sources to capture the sporting, economic, and managerial dimensions of head coach turnover in Serie A.

Match-level outcomes, league standings, and seasonal team performance indicators (e.g., points, wins, draws, losses, goals scored and conceded) are collected from official Serie A records and specialized football statistics databases ([59] and [60]). These sources provide authoritative and consistent information on match results and team performance across seasons. Information on head coaches, including appointment and departure dates, turnover events, nationality, age, and previous career characteristics, is compiled from official club announcements and verified through publicly available archival sources such as club websites, league communications, and reputable football databases ([59, 61], and [60]). This cross-validation procedure is adopted to ensure the accuracy of turnover classifications and the coach characteristics. Team- and player-level indicators are extracted from the Transfermarkt website (see [59]). Furthermore, team value is the aggregate market value of all players in the squad at the start of the season, computed as the sum of individual players' market values as assessed by Transfermarkt. Team market value is commonly used as a proxy for a club's overall sporting and economic strength. In contrast, market value is the estimated transfer value of individual players, based on performance, age, and future potential. These values are expert-based estimates derived from a combination of performance indicators, age, contract duration, and market conditions, and are widely used in the sports economics literature as proxies for economic and sporting strength. Transfermarkt's valuation does not represent an actual transaction price but rather an informed estimate reflecting sporting performance, potential, contractual situation, demand, and market dynamics. Moreover, Transfermarkt market values are expert-based estimates rather than observed transaction prices and therefore should be interpreted as proxies for players' and teams' economic and sporting value rather than exact transfer fees. Finally, these metrics allow us to account for the financial disparity between clubs and the differing levels of expectation placed on coaches managing higher-valued squads. Table 2 summarizes the primary data sources associated with each group. All data sources are merged at the team–season level using team identifiers and season labels. Coach–team spells are constructed by aligning coaching tenures with seasonal calendars, measuring time in days from the start of the season until a turnover event or right-censoring at season's end. When multiple coaches are observed within the same season for a team, each spell is treated as a separate observation. Variables are then systematically checked for internal consistency across sources, and any discrepancies are resolved through manual verification using independent media reports and official club communications. Missing values affect only a small fraction of observations and are handled via listwise deletion. Market values are expressed in euros and aligned to the beginning of each season to ensure temporal consistency and to avoid look-ahead bias. Quantitative variables (including both discrete counts and continuous measures) are treated as numeric predictors. At the same time, categorical attributes are encoded as binary indicators to allow asymmetric effects across categories, which would be obscured by groupwise inclusion. This representation is adopted consistently across all modeling approaches and reflects standard practice in survival analysis and machine-learning applications.

**Table 2** Mapping between variable groups, data sources, and construction notes

| Variable Group | Content | Primary Data Source | Notes |
|---|---|---|---|
| Group 1 | Season timing indicators | Serie A official records; public archival sources | Based on calendar dates and match schedules |
| Group 2 | Coach demographics and career characteristics | Serie A official records; Club announcements; public archival sources | Cross-checked across multiple sources |
| Group 3 | Team and club characteristics | Transfermarkt | Market values used as economic proxies |
| Group 4 | Short-term coach performance | Serie A match data; public archival sources | Computed up to the turnover date |
| Group 5 | Cumulative coach performance | Serie A match data; public archival sources | Computed over entire coaching tenure |
| Group 6 | Seasonal team performance | Serie A match data; public archival sources | Based on full season (38 matches) |

### Variable construction and grouping

The data collected for this study are divided into six groups based on the aspects and characteristics of coaches, teams/squads, and clubs:

- Group 1 contains information on the season;
- Group 2 consists of variables related to the sociodemographic characteristics and information on the previous career of the coach;
- Group 3 includes information on the team and soccer club;
- Group 4 measures the performance of the coach until the turnover;
- Group 5 evaluates the performance of the coach during his entire career with the team;
- Group 6 assesses the team's performance.

Hence, we can measure and capture various aspects of a coach's performance that may influence their career and turnover. In particular, the ratios in group 4 measure the performance in a single season up to the turnover, while the ratios in group 5 allow us to analyze the behavior over the entire career of the coach with that particular team. Thus, it is possible to check whether the turnover is also influenced by team performance in previous seasons. Finally, the ratios in group 6 account for the accomplishments in each season.

In more detail, groups 4 and 5 are expressed as performance ratios and differ in their temporal scope. In fact, variables in group 4 capture short-run performance dynamics up to the turnover event, reflecting recent results and immediate pressure on the coach. They are significantly associated with the hazard of dismissal. On the contrary, variables in group 5 summarize cumulative performance over the entire coach–team spell, capturing long-term effectiveness and sustained managerial quality. Their effects, therefore, reflect persistent performance patterns rather than recent shocks. From a managerial perspective, distinguishing between group 4 and group 5 is crucial. Poor performance in the group 4 ratios indicates recent underperformance and may prompt immediate dismissal decisions driven by short-term results and external pressure. In contrast, unfavourable performance in the group 5 ratios reflects sustained underachievement during the coach's tenure, suggesting a more profound mismatch between the coach and the team's long-term objectives. For example, a coach may exhibit strong cumulative performance over the entire tenure (good ratios of group 5) but experience a sequence of poor recent results (bad ratios in group 4), increasing the short-term risk of dismissal despite overall positive career performance with the team.

Table 3 contains a brief description of all variables caught for all groups, and a detailed description of how the ratios in groups 4–6 are computed is given in Table 10 in the appendix.

### Definition of turnover events

We identify four categories of coach turnovers, as described below:

- *Mutual consent*: the coach has been asked to leave, but the club agrees to classify the departure as a joint decision. In this case, the contract is terminated, and the club does not pay the remaining salary. This can be considered the coach's voluntary exit.
- *Resignation*: the coach decides to stop coaching the squad, and the club does not pay the salary for the remainder of the contract. This is the coach's voluntary decision to leave the club.
- *Sacking*: the coach is fired by the club and continues to be paid (until securing a new contract with another team). Thus, this replacement depends on the club's willingness to pay off the coach's contract.
- *End of contract*: the contract naturally expires and is not renewed.

It should be noted that *End of contract* differs from other exit mechanisms as the expiration date is often predetermined. However, in a competing risks framework, it is treated as a stochastic event because the actual observation of a contract expiration depends on the coach's surviving the competing risks of being sacked or resigning. By including it as a specific exit type rather than treating it as independent censoring, we more accurately reflect the organizational reality where reaching the end of a contract represents a successful completion of the professional relationship.

Using this classification, we are able to analyze both voluntary and involuntary turnovers. The dataset includes both first managerial replacements and caretakers appointed within the same season, resulting in 203 managerial replacements. Figure 1 shows the dynamics of the coach turnovers by season.

Table 4 reports the descriptive statistics of the survival time, expressed in days, according to the type of turnover, where the survival time is defined as the number of days from the start of the season to the turnover date or right-censoring at season's end.

**Table 3** Description of the variables in groups 1–6

| | Label | Short name | Name of variable | Type of variable | Coding/Definition |
|---|---|---|---|---|---|
| **Group 1 (Information on season)** | | | | | |
| 1 | var1.01 | DS | During season | binary | 1 if replacement happens during the season and 0 otherwise |
| 2 | var1.02 | ES | End of season | binary | 1 if replacement happens at the end of the season and 0 otherwise |
| 3 | var1.03 | BFM | Before first match | binary | 1 if replacement happens before the first match and 0 otherwise |
| 4 | var1.04 | FSR | First/second half of season | binary | 1 if replacement happens during the second half of the season and 0 otherwise |
| **Group 2 (Sociodemographic information and previous career of coach)** | | | | | |
| 1 | var2.01 | Age.start | Age of starting coaching | numeric | Age at time of taking over coaching |
| 2 | var2.02 | Age.stop | Age of stopping coaching | numeric | Age at the turnover |
| 3 | var2.03 | Age.first | Age as first coach | numeric | Age at first experience as coach |
| 4 | var2.04 | Age.player | Age of stopping playing | numeric | Age when coach stopped playing |
| 5 | var2.05 | ITA | Italian | binary | 1 if coach is Italian and 0 otherwise |
| 6 | var2.06 | NAT | One nationality | binary | 1 if coach has one nationality and 0 otherwise |
| 7 | var2.07 | POS | Other position in the team | binary | 1 if coach held another position in the coaching team and 0 otherwise |
| 8 | var2.08 | CYT | Coach of youth team | binary | 1 if coach was manager of a youth team and 0 otherwise |
| 9 | var2.09 | EX.P | Ex soccer player | binary | if coach was a soccer player and 0 otherwise |
| 10 | var2.10 | MIDF | Ex soccer player in midfield | binary | 1 if coach played in midfield and 0 otherwise |
| 11 | var2.11 | ATT | Ex soccer player in attack | binary | 1 if coach played in attack and 0 otherwise |
| 12 | var2.12 | DEF | Ex soccer player in defence | binary | 1 if coach played in defence and 0 otherwise |
| 13 | var2.13 | GKeep | Ex soccer player as goalkeeper | binary | 1 if coach was a goalkeeper and 0 otherwise |
| 14 | var2.14 | P&C | Played in the same team that he is coaching | binary | 1 if coach was a player in the team he now coaches and 0 otherwise |
| 15 | var2.15 | FE | Foreign experience | binary | 1 if coach has foreign experience and 0 otherwise |
| 16 | var2.16 | FE.C | Foreign experience as coach | binary | 1 if coach has previously coached abroad and 0 otherwise |
| 17 | var2.17 | FE.P | Foreign experience as player | binary | 1 if coach played abroad and 0 otherwise |
| **Group 3 (Information on team and club)** | | | | | |
| 1 | var3.01 | No.p | # of players in the squad | numeric | |
| 2 | var3.02 | Age.p | Average age of the players | numeric | |
| 3 | var3.03 | No.fp | Number of foreign players in the team | numeric | |
| 4 | var3.04 | Tval | Team value$^{a}$ | numeric | |
| 5 | var3.05 | Mval | Market value$^{b}$ | numeric | |
| 6 | var3.06 | Cup | Football cups | binary | 1 if the team qualified for cups and 0 otherwise |

**Table 3** (continued)

| | | | | | |
|---|---|---|---|---|---|
| **Group 3 (Information on team and club)** | | | | | |
| 7 | var3.07 | Rel | Relegated team | binary | 1 if the team was relegated to a lower division and 0 otherwise |
| **Group 4 (Performance of coach until the turnover)** | | | | | |
| 1 | var4.01 | LM.TMT | Proportion of losses to total matches until the turnover | numeric | |
| 2 | var4.02 | HLM.TMT | Proportion of home losses to total matches until the turnover | numeric | |
| 3 | var4.03 | NM.TMT | Proportion of draws to total matches until the turnover | numeric | |
| 4 | var4.04 | HNM.TMT | Proportion of home draws to total matches until the turnover | numeric | |
| 5 | var4.05 | PPP.TMT | Points per match until the turnover | numeric | |
| 6 | var4.06 | GM.TMT | Goals per match until the turnover | numeric | |
| 7 | var4.07 | HGM.TMT | Home goals per match until the turnover | numeric | |
| 8 | var4.08 | GC.TMT | Goals conceded per match until the turnover | numeric | |
| 9 | var4.09 | HGC.TMT | Home goals conceded per match until the turnover | numeric | |
| **Group 5 (Performance of the coach during his entire career with the team)** | | | | | |
| 1 | var5.01 | LM.TMC | Proportion of losses to total matches coached with the team | numeric | |
| 2 | var5.02 | HLM.TMC | Proportion of home losses to total matches coached with the team | numeric | |
| 3 | var5.03 | NM.TMC | Proportion of draws to total matches coached with the team | numeric | |
| 4 | var5.04 | HNM.TMC | Proportion of home draws to total matches coached with the team | numeric | |
| 5 | var5.05 | PPP.TMC | Points per match over matches coached with the team | numeric | |
| 6 | var5.06 | GM.TMC | Goals per match coached with the team | numeric | |

**Table 3** (continued)

| | | | | |
|---|---|---|---|---|
| **Group 5 (Performance of the coach during his entire career with the team)** | | | | |
| 7 | var5.07 | HGM.TMC | Home goals per match coached with the team | numeric |
| 8 | var5.08 | GC.TMC | Goals conceded per match coached with the team | numeric |
| 9 | var5.09 | HGC.TMC | Home goals conceded per match coached with the team | numeric |
| **Group 6 (Performance of the team)** | | | | |
| 1 | var6.01 | LM.TMS | Proportion of losses to total matches in the season | numeric |
| 2 | var6.02 | HLM.TMS | Proportion of home losses to total matches in the season | numeric |
| 3 | var6.03 | NM.TMS | Proportion of draws to total matches in the season | numeric |
| 4 | var6.04 | HNM.TMS | Proportion of home draws to total matches in the season | numeric |
| 5 | var6.05 | PPP.TMS | Points per match in the season | numeric |
| 6 | var6.06 | GM.TMS | Goals per match in the season | numeric |
| 7 | var6.07 | HGM.TMS | Home goals per match in the season | numeric |
| 8 | var6.08 | GC.TMS | Goals conceded per match in the season | numeric |
| 9 | var6.09 | HGC.TMS | Home goals conceded per match in the season | numeric |

## Empirical results

In this section, we show the main results obtained by our dataset. We identify the most relevant covariates for each competing event using the RSF approach (Sect. "Random Survival Forests"), thereby bypassing the need to impose any parametric or semi-parametric structure on the underlying models.

Then, we identify the most significant covariates using *forward selection* and finally using *LASSO regression*.

After that, we estimate hazard ratios in the model (see Sect. "Cause-specific hazard ratios by exit mechanism") using the variables selected by all three methods.

In more detail, because variables may differ across the three procedures, we include the variables selected by all three methods in the final competing risks model and exclude those not identified by all three variable selection techniques.

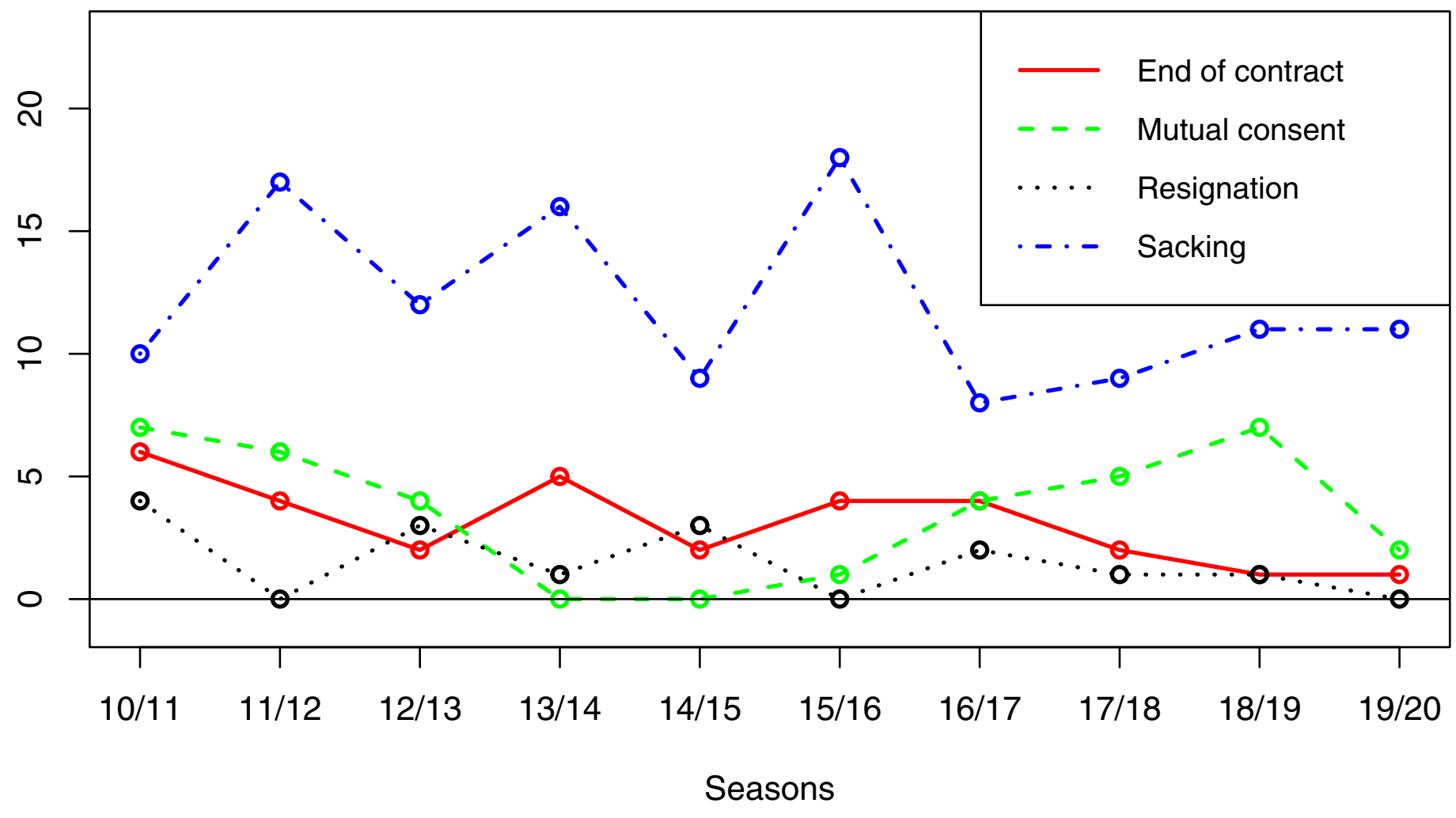

**Fig. 1** Number of coach turnovers broken down by exit for each season in Italian Serie A

**Table 4** Descriptive statistics of survival time, expressed in days, per turnover type

| | # | Mean | S.D. | 25% quartile | Median | 75% quartile |
|---|---|---|---|---|---|---|
| End of contract | 31 | 123.68 | 105.91 | 46.00 | 101.00 | 167.00 |
| Mutual consent | 36 | 214.06 | 117.36 | 109.25 | 207.00 | 331.50 |
| Resignation | 15 | 178.93 | 122.79 | 67.00 | 159.00 | 302.50 |
| Sacking | 121 | 144.27 | 80.88 | 88.00 | 130.00 | 190.00 |
| All | 203 | 212.80 | 121.29 | 108.00 | 196.00 | 364.00 |

It is important to underline that, as suggested by [62], we focus on the cause-specific hazard function of Eq. (4), because the hazard ratio $e^{\beta_{jk}}$ gives the multiplicative effect of covariate $Z_j$ on the instantaneous risk of experiencing event *k*, conditional on the coach still being in charge. This quantity captures how covariates affect the timing of a specific type of turnover among coaches who have not yet experienced any exit. More precisely, this is directly related to the cause-specific hazard, i.e., the instantaneous risk that a turnover of the relevant type occurs at time *t*, given that the coach has not yet experienced any turnover.

In other words, the cause-specific hazard ratio measures the relative change in the instantaneous rate of occurrence of the primary event in subjects who are currently event-free. The rate of occurrence of the event denotes the intensity with which events occur.

### Relevant covariates by exit mechanism

The results of the variable selection methods are presented in Table 5, where we report the number of variables that have been chosen as relevant for all three procedures, for each competing event, and all groups, and in Table 6, where all such variables are listed.

It is worth noting from Tables 5 and 6 that the covariates selected as relevant differ across variable selection procedures and competing events, with only partial overlap among methods. This divergence is expected given the distinct statistical principles underlying RSF, forward selection, and LASSO, and it motivates retaining only variables jointly selected by all three approaches in the final model. Importantly, covariates consistently selected across RSF, LASSO, and forward selection can be interpreted as robust determinants of coach turnover, as their relevance does not depend on a specific estimation or selection framework.

**Table 5** Number of variables per competing event chosen by the three procedures: random survival forest (RSF), forward selection (FS) and LASSO

| | **End of contract** | | | **Mutual Consent** | | | **Resigned** | | | **Sacked** | | |
|---|---|---|---|---|---|---|---|---|---|---|---|---|
| | **RSF** | **FS** | **LASSO** | **RSF** | **FS** | **LASSO** | **RSF** | **FS** | **LASSO** | **RSF** | **FS** | **LASSO** |
| Group 1 | 3 | 2 | 3 | 3 | 2 | 3 | 2 | 2 | 2 | 3 | 3 | 2 |
| Group 2 | 5 | 4 | 3 | 4 | 1 | 4 | 4 | 5 | 4 | 5 | 1 | 4 |
| Group 3 | 2 | 3 | 4 | 4 | 2 | 3 | 5 | 2 | 5 | 2 | 3 | 5 |
| Group 4 | 3 | 1 | 1 | 1 | 5 | 4 | 1 | 5 | 4 | 2 | 5 | 5 |
| Group 5 | 3 | 4 | 6 | 2 | 4 | 5 | 2 | 1 | 1 | 2 | 5 | 4 |
| Group 6 | 5 | 4 | 4 | 1 | 3 | 6 | 3 | 5 | 4 | 3 | 5 | 5 |

#### Cause-specific hazard ratios by exit mechanism

In this section, we show the estimates of the cause-specific hazard functions for all exits associated with coach turnover and assess both the sign and magnitude of covariate effects across competing events. For each group of variables described in Sect. "Data and variable construction" and for each exit type, we estimate the competing risks model including only those covariates jointly selected by all three variable selection procedures (RSF, forward selection, and LASSO) (Table 6). This conservative inclusion rule ensures that the reported effects are robust to different selection criteria and modeling assumptions. Moreover, we report only statistically significant covariates, adopting a significance level of $\alpha = 0.10$.

Furthermore, we focus not only on statistical significance, but also on the substantive interpretation of estimated effects, emphasizing how different dimensions of performance and timing translate into distinct managerial turnover decisions.

For the sake of readability, we clarify how to interpret the estimates of the cause-specific hazard functions. Within the competing risks framework, estimated cause-specific hazard ratios quantify how observable characteristics affect the instantaneous likelihood of a specific type of coach turnover, conditional on the coach remaining in charge. From a managerial perspective, a hazard ratio greater than one indicates factors that accelerate the timing of a coaching replacement and decision to terminate or replace the coach. In contrast, a hazard ratio below one indicates characteristics associated with greater job stability and delayed turnover.

From a practical standpoint, hazard ratios provide a direct indication of how sensitive turnover decisions are to changes in observable performance indicators. For example, hazard ratios substantially greater than 1 imply that relatively small performance deteriorations can lead to rapid increases in dismissal risk. In contrast, hazard ratios well below one indicate factors that provide a stabilizing buffer against turnover. This interpretation allows the estimated effects to be translated into concrete managerial implications for tolerance thresholds, monitoring intensity, and the timing of club board intervention decisions.

A positive coefficient $\beta_{jk} > 0$ (and equivalently, a hazard ratio $\exp(\beta_{jk}) > 1$) implies that an increase in the *j*-th covariate raises the instantaneous risk of exit type *k*, thereby shortening the expected duration of the coach–team spell. In other words, a hazard ratio greater than 1 indicates a covariate that is positively associated with the event probability and thus negatively associated with survival time [63]. For example, a hazard ratio of 1.20 indicates that a one-unit increase in the covariate raises the instantaneous probability of turnover by approximately 20%, conditional on survival up to time *t* [63]. In practical terms, this means that relatively small deteriorations or improvements in performance indicators can substantially alter the timing of managerial decisions, especially in environments such as professional football where performance is continuously monitored. It is important to note that several covariates are defined as proportions or bounded performance measures. Consequently, the reported hazard ratios correspond to a one-unit increase over the entire support of the variable, which represents an extreme and rarely observed change in practice. Large hazard ratios should therefore be interpreted

**Table 6** Selection of variables per competing event given by the three procedures: random survival forest (RSF), forward selection (FS) and LASSO

| | End of contract | | | Mutual Consent | | | Resigned | | | Sacked | | |
|---|---|---|---|---|---|---|---|---|---|---|---|---|
| | **RSF** | **FS** | **LASSO** | **RSF** | **FS** | **LASSO** | **RSF** | **FS** | **LASSO** | **RSF** | **FS** | **LASSO** |
| **Group 1 (Information on season)** | | | | | | | | | | | | |
| DS | ✓ | ✓ | ✓ | ✓ | ✓ | ✓ | | ✓ | | ✓ | ✓ | ✓ |
| ES | ✓ | ✓ | ✓ | ✓ | ✓ | ✓ | ✓ | ✓ | ✓ | ✓ | ✓ | |
| BFM | | | | ✓ | | ✓ | ✓ | | ✓ | | | |
| FSR | ✓ | | ✓ | | | | | | | ✓ | ✓ | ✓ |
| **Group 2 (Sociodemographic information and previous career of coaches)** | | | | | | | | | | | | |
| Age.start | | | | | | | | | | | | |
| Age.stop | ✓ | | | ✓ | | | ✓ | | | ✓ | | ✓ |
| Age.first | | ✓ | ✓ | | | ✓ | | | | | | ✓ |
| Age.player | | | | | | | | | | | | |
| ITA | ✓ | ✓ | ✓ | ✓ | | | ✓ | ✓ | | ✓ | | |
| NAT | | | | | | ✓ | | | | | | ✓ |
| POS | | ✓ | | | | | | ✓ | ✓ | | | |
| CYT | | | | | | | | | | | ✓ | |
| EX.P | | | | | | | | ✓ | | | | |
| MIDF | ✓ | | | | | | | | | ✓ | | ✓ |
| ATT | | | | | | | | | ✓ | | | |
| DEF | | | | | | | | | ✓ | | | |
| GKeep | | | | | ✓ | ✓ | | | | | | |
| P&C | ✓ | ✓ | ✓ | ✓ | | | ✓ | | | ✓ | | |
| FE | ✓ | | | ✓ | | ✓ | ✓ | | | ✓ | | |
| FE.C | | | | | | | | ✓ | ✓ | | | |
| FE.P | | | | | | | | ✓ | | | | |
| **Group 3 (Information on team and club)** | | | | | | | | | | | | |
| No.p | | | | ✓ | | | ✓ | | | | | |
| Age.p | | ✓ | ✓ | ✓ | ✓ | ✓ | ✓ | ✓ | ✓ | | | ✓ |
| No.fp | | | | | | | ✓ | | ✓ | | ✓ | ✓ |

**Table 6** (continued)

| | | | | | | | | | | | | |
|---|---|---|---|---|---|---|---|---|---|---|---|---|
| **Group 3 (Information on team and club)** | | | | | | | | | | | | |
| Tval | | | | ✓ | | | ✓ | | | ✓ | | ✓ |
| Mval | ✓ | ✓ | ✓ | ✓ | | | ✓ | | ✓ | ✓ | | |
| Cup | | | ✓ | | | ✓ | | | ✓ | | ✓ | ✓ |
| Rel | ✓ | ✓ | ✓ | | ✓ | ✓ | | ✓ | ✓ | | ✓ | ✓ |
| **Group 4 (Performance of coach until the turnover)** | | | | | | | | | | | | |
| LM.TMT | | | | | | | | ✓ | ✓ | ✓ | | |
| HLM.TMT | | | | | | | | | | | | ✓ |
| NM.TMT | ✓ | | | | | | ✓ | ✓ | ✓ | | | ✓ |
| HNM.TMT | | | | | ✓ | ✓ | | | | | | |
| PPP.TMT | | ✓ | | | | | | ✓ | ✓ | ✓ | ✓ | ✓ |
| GM.TMT | | | ✓ | | ✓ | ✓ | | | | | ✓ | ✓ |
| HGM.TMT | ✓ | | | ✓ | ✓ | ✓ | | | | | ✓ | ✓ |
| GC.TMT | | | | | ✓ | | | ✓ | | | | |
| HGC.TMT | ✓ | | | | ✓ | ✓ | | ✓ | ✓ | | | |
| **Group 5 (Performance of the coach during his entire career with the team)** | | | | | | | | | | | | |
| LM.TMC | | ✓ | ✓ | | | ✓ | | | | ✓ | ✓ | ✓ |
| HLM.TMC | | | | ✓ | ✓ | ✓ | | | | ✓ | | ✓ |
| NM.TMC | ✓ | | ✓ | | | ✓ | | | | | | |
| HNM.TMC | | ✓ | ✓ | | | | | | | | ✓ | |
| PPP.TMC | ✓ | ✓ | ✓ | ✓ | ✓ | ✓ | ✓ | | | | ✓ | ✓ |
| GM.TMC | | | ✓ | | ✓ | | ✓ | | | | ✓ | |
| HGM.TMC | | | | | | | | | | | ✓ | |
| GC.TMC | | ✓ | | | ✓ | ✓ | | ✓ | ✓ | | | ✓ |
| HGC.TMC | ✓ | | ✓ | | | | | | | | | |
| **Group 6 (Performance of the team)** | | | | | | | | | | | | |
| LM.TMS | ✓ | ✓ | ✓ | ✓ | | | | ✓ | ✓ | | ✓ | ✓ |
| HLM.TMS | | | | | | | | | | | | ✓ |
| NM.TMS | | | | | ✓ | ✓ | ✓ | ✓ | ✓ | ✓ | ✓ | ✓ |

**Table 6** (continued)

| **Group 6 (Performance of the team)** | | | | | | | | | | | | |
|---|---|---|---|---|---|---|---|---|---|---|---|---|
| HNM.TMS | | ✓ | ✓ | | | ✓ | ✓ | | | | ✓ | ✓ |
| PPP.TMS | ✓ | ✓ | ✓ | | | | ✓ | ✓ | ✓ | ✓ | ✓ | ✓ |
| GM.TMS | | ✓ | | | ✓ | ✓ | | | | ✓ | | |
| HGM.TMS | ✓ | | ✓ | | | ✓ | | | | | ✓ | |
| GC.TMS | ✓ | | | | ✓ | ✓ | ✓ | ✓ | | | | |
| HGC.TMS | ✓ | | | | | ✓ | | ✓ | ✓ | | | |

as evidence of strong directional effects rather than as literal marginal impacts of small performance variations. In practical terms, even moderate changes within the observed range of these variables may substantially affect turnover risk, without implying implausibly large instantaneous effects.

Tables 7 and 8 report the estimated coefficients and hazard ratios for the four competing exit mechanisms. In particular, Table 7 presents results separately by groups of variables, whereas Table 8 reports estimates from the specification that includes all relevant covariates simultaneously.

Looking at the results for the first competing event, *end of contract*, Table 7 shows a positive association with ES, P&C, Rel, and PPP.TMT. This suggests that improved team performance and contractual proximity systematically increase the likelihood that a coaching spell ends through non-renewal rather than abrupt termination. In contrast, the adverse effects of LM.TMS and PPP.TMS indicate that sustained medium- and long-term team performance reduces the risk of contract expiration, consistent with longer planning horizons adopted by successful clubs. These results suggest that contract expiration is not merely a passive outcome but often reflects a deliberate strategic choice by clubs to reassess leadership when performance benchmarks are met or marginally exceeded, rather than resorting to disruptive mid-season interventions.

From a managerial perspective, these findings imply that contract non-renewal is used as a strategic adjustment rather than a response to failure, allowing clubs to reassess leadership even in the presence of satisfactory performance, without incurring the costs and disruptions associated with mid-season dismissals.

For *mutual consent* terminations, the hazard is higher toward specific periods of the season (DS and ES) and increases with indicators such as HLM.TMC and PPP.TMC. These results suggest that mutual agreement is more likely when short-term performance deteriorates, and renegotiation becomes a strategic alternative to unilateral dismissal. The negative association with players' average age indicates that experienced squads are less likely to be involved in consensual separations, possibly reflecting greater tactical stability. Then, higher values of HGM.TMT reduce the likelihood of mutual consent, highlighting the stabilizing role of long-term team performance. From a managerial perspective, mutual consent appears to function as a negotiated exit mechanism, activated when short-term performance signals are unfavorable but not sufficiently severe to justify unilateral dismissal, allowing both parties to limit reputational and contractual costs.

In practical terms, this suggests that mutual consent serves as an intermediate governance mechanism, enabling clubs to respond to declining short-term performance while preserving reputational capital and contractual flexibility for both the club and the coach.

*Resignations* are more likely at the end of the season (ES), consistent with voluntary mobility and contract completion. However, the risk of resignation decreases as Age.p, NM.TMT, NM.TMS, and PPP.TMS increase, suggesting that coaches managing stable and well-performing teams face fewer incentives to step down voluntarily. This pattern aligns with theories of career concerns and opportunity costs in managerial labor markets. This pattern supports the interpretation of resignation as a forward-looking career decision, whereby coaches are less inclined to leave voluntarily when managing cohesive, stable teams with sustained performance, reflecting higher exit opportunity costs.

**Table 7** Estimates, standard errors, and the 95% confidence interval of the significant covariates per competing event, estimated separately for each group of variables. Significance levels are shown in parentheses

| | End of contract | | | Mutual Consent | | | Resignation | | | Sacking | | |
|---|---|---|---|---|---|---|---|---|---|---|---|---|
| | **est.** | **s.e.** | **c.i.(95%)** | **est.** | **s.e.** | **c.i.(95%)** | **est.** | **s.e.** | **c.i.(95%)** | **est.** | **s.e.** | **c.i.(95%)** |
| **Group 1 (Information on season)** | | | | | | | | | | | | |
| DS | - | - | - | 3.78 (****) | 0.84 | [2.13; 5.43] | - | - | - | 3.87 (****) | 0.42 | [3.05; 4.70] |
| ES | 2.69 (****) | 0.46 | [1.79; 3.60] | 3.83 (****) | 0.75 | [2.36; 5.31] | 0.98 (****) | 0.81 | [−0.04; 2.00] | | | |
| FSR | - | - | - | - | - | | - | - | - | −1.10 (****) | 0.24 | [−1.58; −0.62] |
| **Group 2 (Sociodemographic information and previous career of coaches)** | | | | | | | | | | | | |
| P&C | 0.85 (**) | 0.38 | [0.11; 1.59] | - | - | - | | - | - | - | - | - |
| **Group 3 (Information on team and club)** | | | | | | | | | | | | |
| Age.p | - | - | - | −0.23 (*) | 0.13 | [−0.49; 0.03] | −0.40 (*) | 0.23 | [−0.85; 0.05] | - | - | - |
| Rel | 1.51 (****) | 0.38 | [0.78; 2.25] | - | - | - | - | - | - | - | - | - |
| **Group 4 (Performance of coach until the turnover)** | | | | | | | | | | | | |
| NM.TMT | - | - | | - | - | - | −6.11 (**) | 3.10 | [−12.18; −0.04] | - | - | - |
| PPP.TMT | - | - | | - | - | - | - | - | - | −2.51 (****) | 0.24 | [−2.97; −2.04] |
| HGM.TMT | - | - | | −3.91 (****) | 0.78 | [5.44; −2.39] | - | - | - | - | - | - |
| **Group 5 (Performance of the coach during his entire career with the team)** | | | | | | | | | | | | |
| LM.TMC | - | - | - | - | - | - | - | - | - | 4.14 (****) | 0.44 | [3.28; 5.00] |
| HLM.TMC | - | - | - | 10.35(****) | 1.64 | [7.13; 13.57] | - | - | - | - | - | - |
| PPP.TMC | 1.45 (****) | 0.33 | [0.81; 2.09] | 1.09 (***) | 0.38 | [0.34; 1.84] | - | - | - | - | - | - |
| **Group 6 (Performance of the team)** | | | | | | | | | | | | |
| LM.TMS | −11.65 (****) | 2.59 | [−16.74; −6.57] | - | - | - | - | - | - | - | - | - |
| NM.TMS | - | - | - | - | - | - | −6.75 (*) | 3.95 | [−14.48; 0.97] | −9.04 (****) | 1.91 | [−12.77; −5.3] |
| PPP.TMS | −5.41 (****) | 1.01 | [−7.39; −3.44] | - | - | - | −1.13 (*) | 0.62 | [−2.36; 0.09] | −3.24 (****) | 0.43 | [−4.08; −2.40] |

– denotes no significance; * denotes that the covariate is significant at the 10% level; ** denotes that the covariate is significant at the 5% level; *** denotes that the covariate is significant at the 1% level; and **** denotes that the covariate is significant at the 0.1% level. DS: *During the season*; ES: *At the end of the season*; FSR: *first/second half of season*; P&C: *Played in the same team he is coaching*; Age.p: *Average age of the players*; Rel: *Relegated team*; NM.TMT: *Proportion of draws to total matches until the turnover*; PPP.TMT: *Points per match until the turnover*; HGM.TMT: *Home goals per match until the turnover*; LM.TMC: *Proportion of losses to total matches coached with the team*; HLM.TMC: *Proportion of home losses to total matches coached with the team*; PPP.TMC: *Points per match coached with the team*; LM.TMS: *Proportion of losses to total matches in the season*; NM.TMS: *Proportion of draws to total matches in the season*; PPP.TMS: *Points per match in the season*

From a managerial standpoint, this pattern indicates that voluntary exits are less driven by immediate performance pressure and more by career incentives, implying that clubs managing stable, competitive teams face a lower risk of unexpected leadership changes.

Finally, the risk of *sacking* is significantly higher during the season (DS), reflecting reactive organizational responses to poor short-term performance. Conversely, the hazard of dismissal decreases in the second half of the season (FSR) and with stronger team performance indicators such as PPP.TMT, NM.TMS, and PPP.TMS. These findings indicate that clubs become more tolerant of underperformance as the season progresses, particularly when longer-term performance signals are favorable. Overall, sacking emerges as the most reactive turnover mechanism, driven primarily by short-term performance pressures. At the same time, clubs appear increasingly tolerant of underperformance as the season progresses when longer-term indicators remain favorable.

These results imply that sacking decisions are predominantly driven by short-term performance shocks, but that clubs implicitly condition such decisions on longer-term performance histories, exhibiting greater tolerance when cumulative indicators remain favorable.

Comparing Tables 7 and 8, some covariates lose statistical significance when all groups are included simultaneously, reflecting correlations across performance dimensions and overlapping information content. Notably, the persistence of key performance indicators–particularly points per match and loss-related measures–across specifications confirms that the main drivers of coach turnover are robust and not an artefact of variable grouping or model specification. Notably, the core findings remain qualitatively unchanged across group-specific models and the full specification, indicating limited sensitivity to alternative variable inclusion choices. Taken together, the results reveal a clear separation between reactive exit mechanisms (such as sacking), which respond to immediate performance shocks, and strategic or negotiated exits (such as contract expiration and resignation), which are shaped by longer-term performance trajectories. Overall, the empirical evidence indicates that clubs differentiate not only between good and bad performance but also between short-term shocks and sustained trends when deciding how and when to replace a head coach.

Taken together, the magnitude and direction of the estimated hazard ratios indicate that clubs systematically differentiate between exit mechanisms depending on both the severity and the persistence of performance signals. Short-term underperformance triggers rapid corrective actions such as sackings, while sustained performance trajectories shape more strategic outcomes such as contract expiration or resignation. These results provide actionable insights into how performance metrics are implicitly weighted in managerial decision-making processes.

To assess multicollinearity among the covariates included in the cause-specific Cox models, we conduct additional diagnostic checks for the models' results shown in Table 8. In particular, variance inflation factors (VIFs) were computed for the final model specifications by exploiting the linear predictor structure of the Cox regression. All estimated VIF values are below the conventional threshold (VIF $< 5$), indicating that multicollinearity is not a serious concern in our analysis (Table 9). Moreover, although the primary objective of this study is explanatory rather than purely predictive, we assessed model adequacy and predictive performance using standard survival-analysis metrics for

**Table 8** Estimates, standard errors, and the 95% confidence interval of the relevant covariates per competing event, estimated by adding all relevant covariates for all groups. Significance levels are shown in parentheses

| | End of contract | | | Mutual Consent | | | Resignation | | | Sacking | | |
|---|---|---|---|---|---|---|---|---|---|---|---|---|
| | **est.** | **s.e.** | **c.i.(95%)** | **est.** | **s.e.** | **c.i.(95%)** | **est.** | **s.e.** | **c.i.(95%)** | **est.** | **s.e.** | **c.i.(95%)** |
| **Group 1 (Information on season)** | | | | | | | | | | | | |
| DS | - | - | - | 2.99 (****) | 0.79 | [1.43; 4.55] | - | - | - | 2.97 (****) | 0.42 | [2.14; 3.80] |
| ES | 2.85 (****) | 0.59 | [1.68; 4.01] | 3.23 (****) | 0.88 | [1.51; 4.95] | 0.67 (-) | 0.54 | [−0.40; 1.74] | - | - | - |
| FSR | - | - | - | - | - | - | - | - | - | −0.06 (-) | 0.26 | [−0.58; 0.46] |
| **Group 2 (Sociodemographic information and previous career of coaches)** | | | | | | | | | | | | |
| P&C | 0.26 (-) | 0.48 | [−0.69; 1.21] | - | - | - | - | - | - | - | - | - |
| **Group 3 (Information on team and club)** | | | | | | | | | | | | |
| Age.p | - | - | - | −0.04 (-) | 0.16 | [−0.37; 0.28] | −0.48 (*) | 0.28 | [−1.02; 0.07] | - | - | - |
| Rel | 0.77 (-) | 0.50 | [−0.21; 1.76] | - | - | - | - | - | - | - | - | - |
| **Group 4 (Performance of coach until the turnover)** | | | | | | | | | | | | |
| NM.TMT | - | - | - | - | - | - | −2.28 (-) | 4.33 | [−10.77; 6.22] | - | - | - |
| PPP.TMT | - | - | - | - | - | - | - | - | - | 1.61 (****) | 0.42 | [0.78; 2.45] |
| HGM.TMT | - | - | - | −2.43 (**) | 0.98 | [−4.34; −0.53] | - | - | - | - | - | - |
| **Group 5 (Performance of the coach during his entire career with the team)** | | | | | | | | | | | | |
| LM.TMC | - | - | - | - | - | - | - | - | 1.46 (***) | 0.69 | [0.49; 2.42] | |
| HLM.TMC | - | - | - | 2.56 (-) | 2.28 | [−1.90; 7.04] | - | - | - | - | - | |
| PPP.TMC | 0.78 (-) | 0.54 | [−0.28; 1.84] | 1.12 (***) | 0.43 | [0.28; 1.98] | - | - | - | - | - | |
| **Group 6 (Performance of the team)** | | | | | | | | | | | | |
| LM.TMS | −16.24 (****) | 3.49 | [−23.08; −9.40] | - | - | - | - | - | - | - | - | |
| NM.TMS | - | - | - | - | - | −4.84 (-) | 6.15 | [−16.90; 7.21] | −4.19 (****) | 0.71 | [−5.60; −2.78] | |
| PPP.TMS | −7.35 (****) | 1.43 | [−10.16; −4.54] | - | - | −1.20 (*) | 0.72 | [−2.62; 0.21] | −5.69 (***) | 2.14 | [−9.90; −1.49] | |

Notes: – denotes no significance; * denotes that the covariate is significant at the 10% level; ** denotes that the covariate is significant at the 5% level; *** denotes that the covariate is significant at the 1% level; and **** denotes that the covariate is significant at the 0.1% level. DS: *During the season*; ES: *At the end of the season*; FSR: *first/second half of season*; P&C: *Played in the same team he is coaching*; Age.p: *Average age of the players*; Rel: *Relegated team*; NM.TMT: *Proportion of draws to total matches until the turnover*; PPP.TMT: *Points per match until the turnover*; HGM.TMT: *Home goals per match until the turnover*; LM.TMC: *Proportion of losses to total matches coached with the team*; HLM.TMC: *Proportion of home losses to total matches coached with the team*; PPP.TMC: *Points per match coached with the team*; LM.TMS: *Proportion of losses to total matches in the season*; NM.TMS: *Proportion of draws to total matches in the season*; PPP.TMS: *Points per match in the season*

the models' results shown in Table 8. For the cause-specific Cox models, goodness-of-fit and discriminatory power are evaluated using the concordance index (C-index), which measures the model's ability to rank survival times correctly. Across competing events, the estimated C-index values indicate satisfactory discriminative performance (Table 8). More specifically, the cause-specific Cox models achieve C-index values between 0.84 and 0.98 across exit types, indicating strong discriminatory ability. Overall, the consistency of selected covariates across methods and the satisfactory predictive performance indicators support the adequacy of the proposed modeling framework. The presence of large hazard ratios for some covariates is primarily driven by variable scaling and event concentration rather than by model misspecification, as confirmed by low multicollinearity diagnostics and stable estimates across alternative specifications.

## Discussion and conclusions

This study analyzes the determinants of head coach turnover in the Italian Serie A over the period 2010–2019 by explicitly modeling multiple exit mechanisms within a competing risks framework. By integrating classical survival analysis with machine-learning-based variable selection techniques, the paper provides a data-driven and comprehensive assessment of both the timing and the nature of coaching replacements. The empirical strategy follows three steps. First, a rich dataset is constructed by merging multiple data sources and distinguishing among four mutually exclusive exit mechanisms: contract expiration, termination by mutual consent, resignation, and sacking. Second, three complementary variable selection procedures–Random Survival Forests, forward selection, and LASSO–are applied to identify the most relevant predictors of coach turnover. As expected, these approaches yield partially distinct sets of covariates, reflecting differences in model structure and selection criteria. To enhance robustness and comparability, the final specification retains only those variables that are consistently selected by all three methods. Third, cause-specific hazard models are estimated to quantify the effects of the selected covariates on each exit mechanism. To mitigate concerns related to model specification and variable inclusion, the empirical strategy deliberately combines three fundamentally different variable selection approaches: Random Survival Forests, forward selection, and LASSO regression. From an applied perspective, this strategy provides practitioners with a stable set of performance indicators that remain informative across alternative modeling choices. These methods rely on distinct assumptions and optimization criteria and therefore identify covariate relevance in complementary ways. By retaining in the final cause-specific hazard models only those predictors consistently selected by all three procedures, the analysis adopts a conservative specification that prioritizes robustness over model complexity. The stability of the main performance-related effects across group-specific and full-model specifications further supports the limited sensitivity of the results to alternative modeling choices. The results highlight the importance of institutional timing in shaping decisions about coach turnover. Turnovers occurring during the season are primarily associated with sackings and mutual-consent exits, indicating that clubs respond rapidly to short-term performance signals when immediate competitive objectives are at stake. By contrast, contract expirations are concentrated at the end of the season, consistent with a more deliberate and strategic evaluation process. These findings indicate that clubs implicitly rely on season-specific decision windows, combining reactive mid-season interventions with strategic end-of-season evaluations. Explicitly recognizing this structure may help clubs to formalize evaluation calendars and to reduce ad hoc decision-making under performance pressure. Performance-related variables emerge as the most robust and consistent determinants of coach turnover across exit types. Short-term performance indicators, such as points per match and loss frequency, emerge as operational early-warning signals that can guide timely intervention decisions. This pattern suggests that clubs integrate both recent outcomes and longer-term performance trajectories when assessing managerial stability. Sustained season-long performance appears to provide implicit protection against dismissal, suggesting the presence of tolerance thresholds in managerial evaluations. In contrast, coaches' personal characteristics and broader team or club attributes exhibit limited explanatory power once performance is accounted for. Although previous studies have documented significant effects of background characteristics, these effects largely vanish when multiple exit mechanisms are modeled separately. This result highlights the importance of distinguishing among heterogeneous turnover processes and illustrates how aggregating exit types may lead to misleading inferences. It also supports the use of transparent, performance-based evaluation criteria in contract renewal and dismissal decisions, potentially reducing ambiguity and discretionary bias.

**Table 9** The VIF of the relevant covariates and the C-index for the competing risks models, estimated by adding all relevant covariates for all groups

| | End of contract | Mutual Consent | Resignation | Sacking |
|---|---|---|---|---|
| | VIF | VIF | VIF | VIF |
| **Group 1 (Information on season)** | | | | |
| DS | - | 4.07 | - | 1.06 |
| ES | 1.59 | 4.23 | 1.05 | - |
| FSR | - | - | - | 1.17 |
| **Group 2 (Sociodemographic information and previous career of coaches)** | | | | |
| P&C | 1.24 | - | - | - |
| **Group 3 (Information on team and club)** | | | | |
| Age.p | - | 1.07 | 1.04 | - |
| Rel | 1.65 | - | - | - |
| **Group 4 (Performance of coach until the turnover)** | | | | |
| NM.TMT | - | - | 2.50 | - |
| PPP.TMT | - | - | - | 2.48 |
| HGM.TMT | - | 1.58 | - | - |
| **Group 5 (Performance of the coach during his entire career with the team)** | | | | |
| LM.TMC | - | - | - | 1.18 |
| HLM.TMC | - | 1.50 | - | - |
| PPP.TMC | 2.34 | 1.16 | - | - |
| **Group 6 (Performance of the team)** | | | | |
| LM.TMS | 2.02 | - | - | - |
| NM.TMS | - | - | 1.63 | 1.44 |
| PPP.TMS | 1.47 | - | 3.24 | 2.84 |
| C-index (s.e.) | 0.98 (0.006) | 0.84 (0.039) | 0.86 (0.039) | 0.94 (0.090) |

Comparisons between group-specific specifications and the full model reveal that some covariates lose statistical significance when multiple performance dimensions are included simultaneously. This outcome reflects correlations among short-term, cumulative, and season-long performance indicators rather than model instability. Within this context, some coefficients (such as the short-term performance indicator PPP.TMT) exhibit sign changes when moving from group-specific specifications to the full model. These patterns should not be interpreted as contradictory evidence, but rather as conditional effects arising from the simultaneous inclusion of multiple performance dimensions. Once cumulative and season-level performance measures are controlled for, short-term indicators capture deviations from an established performance trajectory. In managerial terms, this implies that clubs may proceed with dismissal decisions despite temporary performance improvements when longer-term outcomes remain unsatisfactory. Such behavior is consistent with delayed or threshold-based decision-making in professional football. It aligns with evidence that clubs integrate performance information over multiple time horizons rather than reacting solely to recent results. Overall, the consistency of selected covariates across methods and the satisfactory predictive performance indicators support the adequacy of the proposed modeling framework for both explanatory and applied decision-making purposes. Furthermore, the analysis shows that observable performance metrics–particularly points per match and cumulative losses–constitute the primary drivers of coach turnover in Serie A, while personal characteristics play a secondary role. By explicitly accounting for competing exit mechanisms and leveraging multiple variable selection techniques, this study extends the existing literature and provides a more nuanced understanding of managerial turnover in professional football. Despite its contributions, the study has some limitations. A limitation concerns external validity, as the empirical analysis focuses exclusively on the Italian Serie A. This choice is intentional and intended to preserve institutional homogeneity in league structure, competitive format, and governance mechanisms. While turnover dynamics may differ across national leagues due to regulatory, cultural, and organizational factors, restricting attention to a single top-tier league allows for more precise identification of performance-related effects within a consistent institutional environment. Extending the analysis to multiple leagues and conducting cross-country comparisons represents a valuable direction for future research, particularly to assess the generalizability of the competing risks framework across different football labor markets. Another limitation is that the analysis considers only single coaching spells. Future research could incorporate multiple spells or multi-state models to capture more complex career trajectories. Moreover, the covariates are treated as time-invariant. This is a deliberate modeling choice, as the variables capture cumulative or average performance information rather than high-frequency match-level dynamics. While time-varying covariates could provide additional insights into short-term performance shocks, their inclusion would require more granular data and a different empirical framework. Exploring time-varying specifications represents a natural extension of this work. These extensions represent promising directions for future research.

### Managerial and practical implications

The findings provide several insights relevant to decision-making in professional football organizations. Season timing plays a central role in shaping decisions about coach turnover, with distinct patterns emerging across exit mechanisms. Sackings and mutual-consent terminations are substantially more likely during the season, indicating that clubs tend to react swiftly to short-term performance deterioration when competitive objectives are immediately at risk. In contrast, contract expirations are predominantly concentrated at the end of the season, reflecting a more structured and strategic evaluation process. This evidence suggests that clubs operate under a dual-decision framework that combines short-term corrective interventions with longer-term strategic assessments. Performance indicators represent the most informative and operational predictors of managerial stability. Variables such as points per match and the proportion of losses strongly affect dismissal risk, confirming that observable on-field performance remains the primary basis for turnover decisions. From a practical perspective, these results indicate that continuously updated performance metrics can serve as transparent benchmarks for monitoring managerial risk, enabling clubs to identify early warning signals of potential turnover. Home performance emerges as a particularly relevant dimension in negotiated exits. The negative association between goals scored in home matches and the probability of turnover by mutual consent suggests that strong home performances may reduce external pressure from supporters and local media. As a result, clubs appear less inclined to pursue consensual separations when fan-facing outcomes remain favorable, even if overall performance is mixed. This finding highlights the role of reputational and contextual factors in shaping managerial stability. Cumulative performance over a coach's entire tenure plays a crucial role in sacking decisions. Rather than responding exclusively to isolated performance shocks, clubs appear to evaluate longer-term performance trajectories, with poor cumulative results progressively eroding managerial trust. Conversely, a history of solid performance provides protection against immediate dismissal during temporary downturns. In contrast, coaches' personal characteristics (such as age, nationality, or playing background) have limited explanatory power once performance measures are controlled for. This suggests that, within the Italian Serie A, turnover decisions are predominantly outcome-oriented and aligned with merit-based evaluation principles. Taken together, these results indicate that coach turnover decisions reflect a complex interaction between short-term performance signals, cumulative performance histories, and institutional decision windows. By distinguishing among exit mechanisms and adopting a data-driven modeling strategy, the analysis offers actionable insights for clubs seeking to balance accountability, contractual stability, and organizational continuity.

## Appendix A

In Table 10, we describe how the ratios in groups 4–6 are computed. In particular, the column "Definition of variable" lists the names of items we collect to compute them.

**Table 10** Detailed description of how the ratios in groups 4–6 are computed

| | Label | Short name | Name of variable | Definition of variable |
|---|---|---|---|---|
| Group 4 (Performance of coach until the turnover) | | | | |
| 1 | var4.01 | LM.TMT | Proportion of losses to total matches until the turnover | $\frac{\text{Lost matches}}{\text{Total matches coached until the turnover}}$ |
| 2 | var4.02 | HLM.TMT | Proportion of home losses to total matches until the turnover | $\frac{\text{Home lost matches}}{\text{Total matches coached until the turnover}}$ |
| 3 | var4.03 | NM.TMT | Proportion of draws to total matches until the turnover | $\frac{\text{Drawn matches}}{\text{Total matches coached until the turnover}}$ |
| 4 | var4.04 | HNM.TMT | Proportion of home draws to total matches until the turnover | $\frac{\text{Home drawn matches}}{\text{Total matches coached until the turnover}}$ |
| 5 | var4.05 | PPP.TMT | Points per match until the turnover | $\frac{3*\text{Won matches}+1*\text{Drawn matches}}{\text{Total matches coached until the turnover}}$ |
| 6 | var4.06 | GM.TMT | Goals per match until the turnover | $\frac{\text{Goals scored}}{\text{Total matches coached until the turnover}}$ |
| 7 | var4.07 | HGM.TMT | Home goals per match until the turnover | $\frac{\text{Home goals scored}}{\text{Total matches coached until the turnover}}$ |
| 8 | var4.08 | GC.TMT | Goals conceded per match until the turnover | $\frac{\text{Goals conceded}}{\text{Total matches coached until the turnover}}$ |
| 9 | var4.09 | HGC.TMT | Home goals conceded per match until the turnover | $\frac{\text{Home goals conceded}}{\text{Total matches coached until the turnover}}$ |
| **Group 5 (Performance of the coach during his entire career with the team)** | | | | |
| 1 | var5.01 | LM.TMC | Proportion of losses to total matches coached with the team | $\frac{\text{Lost matches}}{\text{Total matches coached with the team}}$ |
| 2 | var5.02 | HLM.TMC | Proportion of home losses to total matches coached with the team | $\frac{\text{Home lost matches}}{\text{Total matches coached with the team}}$ |
| 3 | var5.03 | NM.TMC | Proportion of draws to total matches coached with the team | $\frac{\text{Drawn matches}}{\text{Total matches coached with the team}}$ |
| 4 | var5.04 | HNM.TMC | Proportion of home draws to total matches coached with the team | $\frac{\text{Home drawn matches}}{\text{Total matches coached with the team}}$ |
| 5 | var5.05 | PPP.TMC | Points per match coached with the team | $\frac{3*\text{Won matches}+1*\text{Drawn matches}}{\text{Total matches coached with the team}}$ |
| 6 | var5.06 | GM.TMC | Goals per match coached with the team | $\frac{\text{Goals scored}}{\text{Total matches coached with the team}}$ |
| 7 | var5.07 | HGM.TMC | Home goals per match coached with the team | $\frac{\text{Home goals scored}}{\text{Total matches coached with the team}}$ |
| 8 | var5.08 | GC.TMC | Goals conceded per match coached with the team | $\frac{\text{Goals conceded}}{\text{Total matches coached with the team}}$ |
| 9 | var5.09 | HGC.TMC | Home goals conceded per match coached with the team | $\frac{\text{Home goals conceded}}{\text{Total matches coached with the team}}$ |
| **Group 6 (Performance of the team)** | | | | |
| 1 | var6.01 | LM.TMS | Proportion of losses to total matches in the season | $\frac{\text{Lost matches}}{\text{Total matches played in season}}$ |
| 2 | var6.02 | HLM.TMS | Proportion of home losses to total matches in the season | $\frac{\text{Home lost matches}}{\text{Total matches played in season}}$ |
| 3 | var6.03 | NM.TMS | Proportion of draws to total matches in the season | $\frac{\text{Drawn matches}}{\text{Total matches played in season}}$ |
| 4 | var6.04 | HNM.TMS | Proportion of home draws to total matches in the season | $\frac{\text{Home drawn matches}}{\text{Total matches played in season}}$ |
| 5 | var6.05 | PPP.TMS | Points per match in the season | $\frac{3*\text{Won matches}+1*\text{Drawn matches}}{\text{Total matches played in season}}$ |
| 6 | var6.06 | GM.TMS | Goals per match in the season | $\frac{\text{Goals scored}}{\text{Total matches played in season}}$ |
| 7 | var6.07 | HGM.TMS | Home goals per match in the season | $\frac{\text{Home goals scored}}{\text{Total matches played in season}}$ |
| 8 | var6.08 | GC.TMS | Goals conceded per match in the season | $\frac{\text{Goals conceded}}{\text{Total matches played in season}}$ |
| 9 | var6.09 | HGC.TMS | Home goals conceded per match in the season | $\frac{\text{Home goals conceded}}{\text{Total matches played in season}}$ |

**Author contributions**

M.R. conceived the study and was responsible for data curation. M.R. and M.Z. conceptualized the methodological framework and developed the software code. All authors analyzed the data and discussed the results. All authors contributed to writing, reviewing, and editing the manuscript. All authors approved the final version of the manuscript.

**Funding**

For the third and fourth authors, this work was partially supported by project PID2023-149087NB-I00, MICIU/AEI/10.13039/501100011033, funded by the Spanish Ministry of Science and Innovation and co-funded by the FEDER programme. The fourth author also acknowledges financial support from the IMAG–María de Maeztu grant CEX2020-001105-M, AEI/10.13039/501100011033.

**Data availability**

The data that support the findings of this study are available from the corresponding author upon reasonable request.

## Declarations

**Competing interests**

The authors declare no competing interests.

## Publisher's Note